\documentclass[twocolumn,showpacs,preprintnumbers,amsmath,amssymb]{revtex4}

\usepackage{graphicx}
\usepackage{dcolumn}
\usepackage{bm}

\begin{document}

\preprint{APS/123-QED}

\title{Pressure-induced ferromagnetism with strong Ising-type anisotropy in YbCu$_2$Si$_2$\footnote{Phys. Rev. B {\bf 89}, 035127 (2014).}}

\author{Naoyuki Tateiwa$^{1}$}
 \email{tateiwa.naoyuki@jaea.go.jp} 
\author{Tatsuma D. Matsuda$^{1,2}$}%
\author{Yoshinori Haga$^{1}$}%
\author{Zachary Fisk$^{1,3}$}

\affiliation{
$^{1}$Advanced Science Research Center, Japan Atomic Energy Agency, Tokai, Naka, Ibaraki 319-1195, Japan\\
$^2$Department of Physics, Tokyo Metropolitan University, Hachioji, Tokyo 192-0397, Japan\\
$^{3}$University of California, Irvine, California 92697, USA\\
}
\date{\today}

\begin{abstract}
We report dc magnetic measurements on YbCu$_2$Si$_2$ at pressures above 10 GPa using a miniature ceramic anvil cell. YbCu$_2$Si$_2$ shows a pressure-induced transition from the non-magnetic to a magnetic phase at 8 GPa. We find a spontaneous dc magnetization in the pressure-induced phase above 9.4 GPa. The pressure dependence of the ferromagnetic transition temperature $T_{\rm C}$ and the spontaneous magnetic moment ${\mu}_0$ at 2.0 K have been determined. The value of ${\mu}_0$ in the present macroscopic measurement is less than half of that determined via M\"{o}ssbauer experiment. The difference may be attributed to spatial phase separation between the ferromagnetic and paramagnetic phases. This separation suggests that the pressure-induced phase boundary between the paramagnetic and ferromagnetic states is of first order. Further, we have studied the magnetic anisotropy in the pressure-induced ferromagnetic state. The effect of pressure on the magnetization with magnetic field along the magnetic easy $c$-axis is much larger than for field along the hard $a$-axis in the tetragonal structure. The pressure-induced phase has strong Ising-type uniaxial anisotropy, consistent with the two crystal electric field (CEF) models proposed for YbCu$_2$Si$_2$. 

\end{abstract}


\maketitle

\section{Introduction}
In recent years there has been growing interest in strongly correlated electron systems of rare earth and actinide compounds located at  or close to a magnetic quantum critical point (QCP)\cite{flouquet}. The electronic state of such systems can often be tuned with pressure or magnetic field. Unconventional superconductivity and non-Fermi liquid behavior have been observed near pressure-induced magnetic to non-magnetic phase boundaries in many cerium compounds such as CeIn$_3$\cite{mathur}. The novel physical phenomena have been studied from the view point of the quantum criticality. Such phenomena might be expected in ytterbium compounds since the Yb is considered to be a ``hole" equivalent of Ce. Indeed, anomalous physical properties have been reported and extensively studied in YbRh$_2$Si$_2$ and $\beta$-YbAlB$_4$\cite{gegenwart,matsumoto}.

Application of pressure tends to drive the Yb ion from nonmagnetic Yb$^{2+}$ ($4f^{14}$) to magnetic Yb$^{3+}$ ($4f^{13}$) states. A magnetic ordered state is stabilized at higher pressures. A pressure-induced magnetic phase has been reported in a number of Yb compounds. In most cases the pressure-induced magnetic phase has been detected via ac magnetic susceptibility measurements. There have been few studies of detailed magnetic properties of a pressure-induced phase using dc magnetization measurements. This is due to the common experimental constraint that the maximum pressure is only 1.5 GPa for the most commonly used piston cylinder type cell in a commercial superconducting quantum interference device (SQUID)\cite{kamarad}. 
  
 Recently, we have developed a miniature ceramic anvil cell (mCAC) for magnetic measurements at pressures above 10 GPa with use of the SQUID magnetometer\cite{tateiwa1,tateiwa2,tateiwa3}. Thanks to the simplicity of cell structure, the mCAC can detect the ferromagnetic ordered state whose spontaneous magnetic moment is significantly less than 1.0 ${\mu}_{\rm B}$ per a magnetic ion. The cell enables us to make a quantitative study of the pressure-induced phases in Yb compounds. We report here a study of the anisotropic magnetic properties of the pressure-induced phase in YbCu$_2$Si$_2$. 
 
  YbCu$_2$Si$_2$ crystallizes in the tetragonal ThCr$_2$Si$_2$- structure. This is a paramagnetic compound with a moderately high value of the linear specific heat coefficient ${\gamma}{\,}{\simeq}$ 135 mJK$^{-2}$mol${^{-1}}$\cite{dung,matsuda1}. Previous high pressure studies suggested a pressure-induced, possibly ferromagnetic ordered state above 8 GPa from ac magnetic susceptibility measurements and M\"{o}ssbauer experiment\cite{yadri,winkelmann,colombier,fernandez1}. It is therefore important to detect the ferromagnetic component from dc magnetic measurement at high pressure. In this study we have measured the magnetization of YbCu$_2$Si$_2$ with our mCAC.  

  \section{Experimental}
   Single crystals of YbCu$_2$Si$_2$ were grown from Sn flux\cite{dung,matsuda1}. We have used our miniature ceramic-anvil high-pressure cell mCAC with the 0.6 mm culet anvils\cite{tateiwa1,tateiwa2,tateiwa3}. The Cu-Be gasket was preindented to 0.08 mm from the initial thickness of 0.30 mm. The diameter of the sample space in the gasket was 0.20 mm. To study the anisotropy of the magnetic properties in YbCu$_2$Si$_2$, two single crystals were measured with magnetic field applied parallel to the magnetic easy $c$-axis (the [001] direction) and the hard $a$-axis ([100] direction) in the tetragonal crystal structure. The sizes of the single crystal samples were 0.11 $\times$ 0.09 $\times$ 0.03 mm$^3$ and 0.10 $\times$ 0.09 $\times$ 0.02 mm$^3$ for magnetic measurements with magnetic field along the $c$-axis and the $a$-axis, respectively. The sample and a Pb pressure sensor were placed in the sample space filled with glycerin as pressure-transmitting medium\cite{tateiwa4}. The pressure values at low temperatures were determined by the pressure dependence of the superconducting transition temperature of Pb\cite{smith,eiling,wittig}. The pressure medium glycerin solidifies at 5 GPa at room temperature. The pressure inhomogeneity was estimated as ${\Delta}P$ $\sim$ 1 GPa above  10 GPa. The demagnetization effect needs to be taken into account in the pressure-induced ferromagnetic state. The internal field values $H_{int}$ were determined by subtracting the demagnetizing field given by $H_{int}$ = $H_{appl}$ - $DM$. Here,  $H_{appl}$ is the external magnetic field and $D$ is the demagnetizing factor. Error bars in Figure 1 (b) indicate possible errors in the estimation of the magnetization.
      
     \begin{figure}[t]
\includegraphics[width=6.7cm]{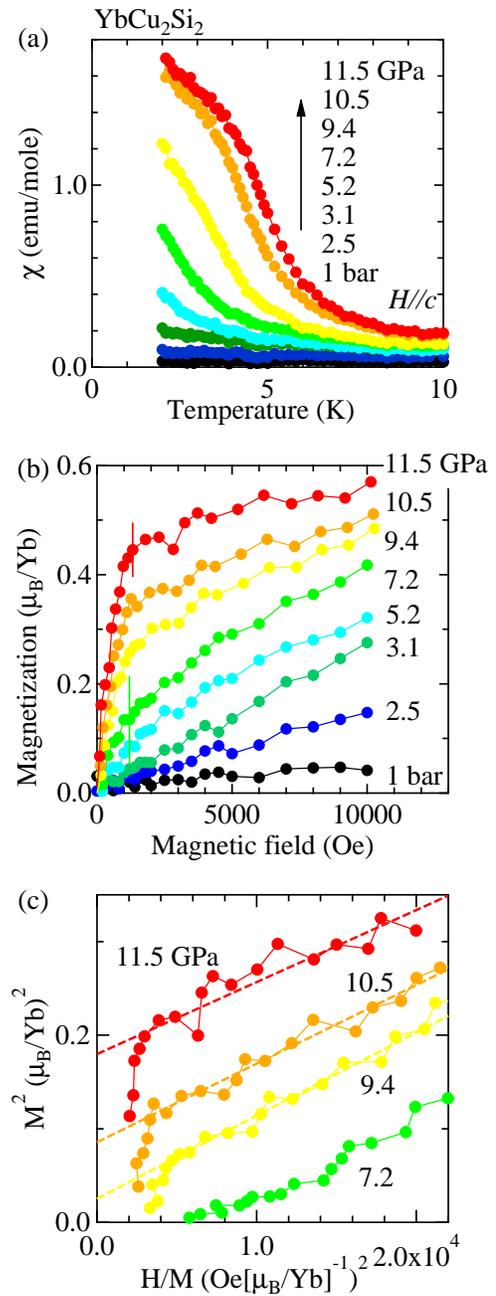}

\caption{\label{fig:epsart}(Color online)(a) Temperature dependence of the magnetic susceptibility ${\chi}$ under magnetic field of 1 kOe applied along the magnetic easy $c$-axis ($H$ $||$ [001]). (b) magnetic field dependence of the magnetization and (c) Arrott-plots of the magnetization measured at 2.0 K and at 1 bar, 2.5, 3.1, 5.2, 7.2, 9.4, 10.5, and 11.5 GPa. }
\end{figure}

  \section{Results and Discussions}
 Figure 1 shows (a) temperature dependence of the magnetic susceptibility ${\chi}$ in a magnetic field of 1 kOe applied along the magnetic easy $c$-axis ($H$ $||$ [001]) and (b) magnetic field dependence of the magnetization measured at 2.0 K and at 1 bar, 2.5, 3.1, 5.2, 7.2, 9.4, 10.5, and 11.5 GPa. At 1 bar, ${\chi}$ shows an almost temperature independent value of ${\chi}$  = 0.03 emu/mole below 10 K and the magnetization increases linearly with increasing magnetic field, consistent with the previous study\cite{dung}. Application of pressure above 5 GPa induces a low temperature upturn in ${\chi}$ and a non-linear increase of the magnetization in low fields. At 9.4, 10.5, and 11.5 GPa, the magnetization shows typical ferromagnetic behavior with the magnetic susceptibility, diverging at low temperatures. These results indicate that the pressure-induced magnetic transition in YbCu$_2$Si$_2$ is ferromagnetic.
    \begin{figure}[t]
\includegraphics[width=6.5cm]{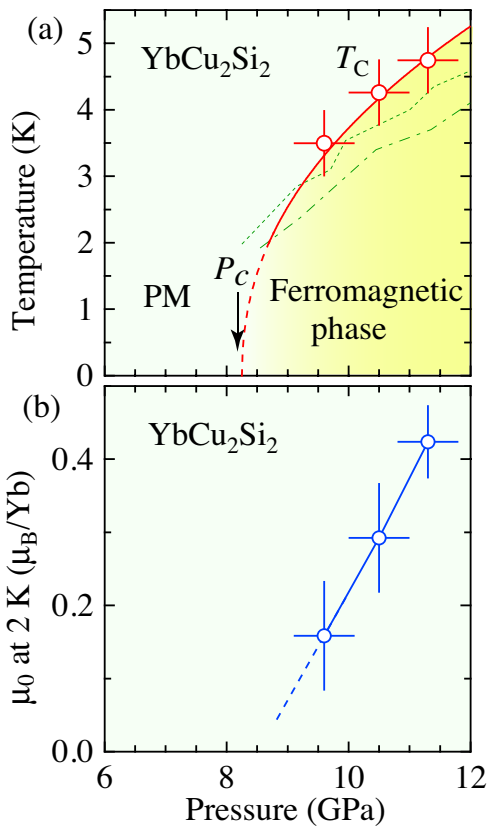}

\caption{\label{fig:epsart}(Color online)(a)Temperature-pressure phase diagram of YbCu$_2$Si$_2$. Circles indicate the ferromagnetic transition temperature $T_{\rm C}$. Dotted and dashed-dotted lines indicate the pressure dependences of $T_{\rm C}$ for the sample with RRR = 200 in the previous study\cite{fernandez1}. The former and the latter lines were determined by the ac magnetic susceptibility measurement and the ac calorimetry, respectively. (b) Pressure dependence of the spontaneous magnetic moment ${\mu}_0$ determined at 2.0 K. }
\end{figure}

  The spontaneous magnetic moment ${\mu}_0$ is determined above 9.4 GPa from the Arrott-plot shown in Fig.1 (c). The values of ${\mu}_0$ at 2.0 K are estimated as 0.16 $\pm$ 0.08, 0.30 $\pm$ 0.08, and 0.42 $\pm$ 0.05 ${\mu}_{\rm B}$/Yb at 9.4, 10.2, and 11.5 GPa, respectively. The ferromagnetic transition temperatures $T_{\rm C}$ at 9.4, 10.5, and 11.5 GPa are estimated as 3.5 $\pm$ 0.5, 4.3 $\pm$ 0.5, and 4.7 $\pm$ 0.5 K, respectively, from the peak position in the temperature derivative of the magnetic susceptibility ${\partial {\chi}}/{\partial T}$. 

 Figure 2 shows the pressure dependences of (a) the ferromagnetic transition temperature $T_{\rm C}$ and (b) the spontaneous magnetic moment ${\mu}_0$ at 2.0 K in YbCu$_2$Si$_2$.  A ferromagnetic transition was not observed down to 2.0 K at 8.8 GPa (data not shown). The transition may occur below 2.0 K. The critical pressure $P_c$ for the ferromagnetic state may be located between 8.0 and 8.5 GPa. Reference 17 reported that the pressure effect on $T_{\rm C}$ depends largely on the sample quality\cite{fernandez1}. The present pressure dependence of $T_{\rm C}$ is consistent with those for samples with the similar quality (RRR = 200) in the previous study, shown as dotted and dashed-dotted lines in Fig. 2 (a). The former and the latter lines were determined by the ac magnetic susceptibility measurement and the ac calorimetry, respectively. 

It has been established that, above critical pressure $P_c$, the transition to the ferromagnetic phase in YbCu$_2$Si$_2$ is of first order\cite{colombier,fernandez1}. Indeed, the ac magnetic susceptibility measurement showed a sudden appearance of the ferromagnetic transition above 1 K\cite{fernandez1}.  However, no sharp anomaly at $T_{\rm C}$ is observed in the temperature dependence of the magnetization at any pressure above 9.4 GPa, indicating a second order phase transition. We suggest that the ferromagnetic transition changes from the first to the second order phase transition at a somewhat higher pressure than $P_c$ in YbCu$_2$Si$_2$. $P_c$ may be a weakly first-order critical point. This may be a reason for absence of the Non-Fermi liquid behavior in the resistivity $\rho$. It shows a typical Fermi liquid behavior ${\rho}={{\rho}_0}+AT^2$ down to 30 mK around $P_c$, where ${\rho}_0$ is the residual resistivity\cite{yadri}. The value of $A$ increases continuously as a function of the pressure but it does not show a divergent behavior around $P_c$. Several ferromagnets such as ZrZn$_2$\cite{uhlarz}, Co(S$_{1-x}$Se$_x$)$_2$\cite{goto}, MnSi\cite{thessieu} and UGe$_2$\cite{pfleiderer} have a tricritical point where the paramagnet to ferromagnet transition changes from a second-order to a first order phase transition when driven toward the QCP by applying external pressure or chemical pressure. This seems to be a general property of ferromagnets as has been theoretically discussed\cite{belitz}. 

We discuss the pressure-induced ferromagnetism in YbCu$_2$Si$_2$ from two points of views. There are two crystal electric field (CEF) models (I and I') proposed for YbCu$_2$Si$_2$ in previous studies\cite{dung}. The values of the magnetic moment expected from the doublet ground state are 2.70 and 2.29 ${\mu}_{\rm B}$/Yb for the CEF models I and I', respectively. We compare the values in the CEF models with that determined in the M\"{o}ssbauer experiment (i) and that determined in the present macroscopic measurement (ii). 

(i) The values of the magnetic moment in the CEF models are more than two times larger than that (1.25 ${\mu}_{\rm B}$/Yb) determined with the M\"{o}ssbauer experiment at 8.9 GPa at 1.8 K\cite{winkelmann}. The reduced magnetic moment in the M\"{o}ssbauer experiment may be due to the Kondo effect. Resonant inelastic x-ray scattering measurement showed that the value of the Yb valence is 2.88 at 7 K near $P_c$\cite{fernandez2}. Thus, ferromagnetism appears in the mixed valence state in YbCu$_2$Si$_2$. Contrary to cerium compounds, a magnetic ordering can appear in the intermediate valence state (${n_{4f}}$ $\ll$ 1) of the Yb systems, where ${n_{4f}}$ is the occupation number of the $4f$ level\cite{harima,flouquet2,miyake}. Differences in the magnetic properties between the Ce and Yb systems arise from differing hierarchies of the energy scales of the Kondo temperature $T_{\rm K}$, the $4f$-band width ${\Delta}_{4f}$ and the splitting energy between ground and first excited states in the CEF levels ${\Delta}_{CEF}$\cite{harima,flouquet2,miyake}. $T_{\rm K}$ of Yb systems could be smaller or comparable to ${\Delta}_{CEF}$ because of the smaller ${\Delta}_{4f}$ in Yb systems than that in cerium systems. In YbCu$_2$Si$_2$, the electrical resistivity under high pressure suggests that $T_{\rm K}$ is less than 50 K at around $P_c$\cite{yadri}. The value of $T_{\rm K}$ is lower than that of ${\Delta}_{CEF}$ in the model I and I'\cite{dung}. The linear specific heat  coefficient $\gamma$ is estimated as $\gamma$ $\sim$ 1 J/mol$\cdot$K$^2$ at $P_c$ from the coefficient $A$ of the $T^2$-term in the resistivity with the Kadowaki-Woods relation\cite{yadri,kadowaki}. The pressure-induced phase in YbCu$_2$Si$_2$ is a ferromagnetic heavy fermion system with the intermediate valence of the Yb ion. This is opposed to Ce-systems where the magnetic ordering or heavy fermium states are usually restricted to the trivalent configuration (${n_{4f}}$ $\sim$ 1.0)\cite{harima,flouquet2,miyake}. 

     \begin{figure}[t]
\includegraphics[width=7cm]{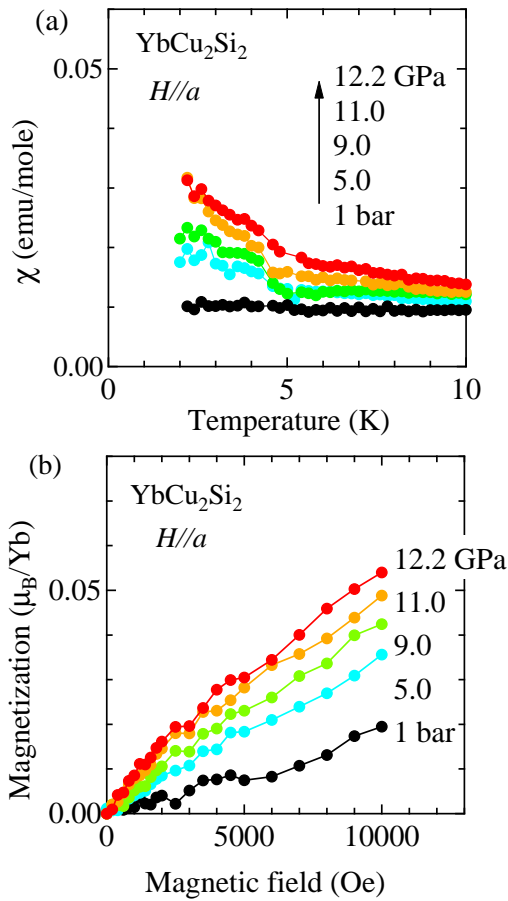}

\caption{\label{fig:epsart}(Color online)(a)Temperature dependence of the magnetic susceptibility ${\chi}$ under magnetic field of 1 kOe applied along the magnetic hard $a$-axis ($H$ $||$ [100]) and (b) magnetic field dependence of the magnetization measured at 2.0 K and at 1 bar, 5.0, 9.0, 11.0, and 12.2 GPa.}
\end{figure}    

 (ii) In the present macroscopic magnetic measurement, the spontaneous magnetic moment ${\mu}_0$ at 2.0 K and 9.4 GPa is less than half of that with the M\"{o}ssbauer experiment\cite{winkelmann}. This difference may be due to a spatial phase separation between the paramagnetic and ferromagnetic states suggested in the M\"{o}ssbauer spectrum. The value of ${\mu}_0$ increases with increasing pressure. The continuous change in ${\mu}_0$ around the critical pressure $P_c$ is difficult to understand because the pressure induced change is of first order\cite{colombier,fernandez1}. We point out two possibilities. One is that the volume fraction of the ferromagnetic phase increases as a function of pressure and the other is that the pressure change of ${\mu}_0$ reflects the increase of the Yb valence above $P_c$ seen in the resonant inelastic x-ray scattering measurement\cite{fernandez2}. The phase separation suggests a first order phase boundary between the paramagnetic and the ferromagnetic phases.

 Figure 3 shows (a) temperature dependence of the magnetic susceptibility ${\chi}$ in magnetic field of 1 kOe applied along the magnetic hard $a$-axis ($H$ $||$ [100]) and (b) magnetic field dependence of the magnetization measured at 2.0 K and at 1 bar, 5.0, 9.0, 11.0, and 12.2 GPa.  Compared with the magnetization data for $H$ $||$ $c$, the pressure effect on the magnetization for $H$ $||$ $a$ is significantly smaller. The value of ${\chi}$ at 2.0 K is increased from 0.01 emu/mole at 1 bar to 0.03 emu/mole at 12.2 GPa. The magnetization curve does not show a ferromagnetic behavior at higher pressures. The magnetic field induced moment at 10 kOe is 0.038 ${\mu}_{\rm B}$/Yb at 12.2 GPa, one order of magnitude smaller than that (0.57 ${\mu}_{\rm B}$/Yb) with magnetic field applied along the easy $c$-axis at 11.5 GPa. The pressure-induced ferromagnetic phase has strong uniaxial anisotropy. The Ising character of the magnetic property is suggested from the two CEF models proposed for YbCu$_2$Si$_2$\cite{dung}. 

The Ising-type magnetic fluctuation can induce the spin-triplet $p$-wave superconductivity around the ferromagnetic QCP\cite{flouquet}.  A motivation for the previous high pressure studies on YbCu$_2$Si$_2$ was to search for the superconductivity around $P_c$. However, the superconductivity has not been found in resistivity measurements down to 30 mK\cite{yadri}. Theoretically, the superconducting transition temperature for spin triplet $p$-wave pairing around the ferromagnetic QCP is largely lower than that of the spin singlet $d$-wave superconductivity around the antiferromagnetic QCP\cite{wang}. In Ce systems, CeIn$_3$ and CeRhIn$_5$ exhibit superconductivity under high pressure where $T_{sc}$ attains maximum values of 0.2 K and 2.2 K, respectively\cite{mathur,hegger}. In uranium systems, the superconductivity appears in the ferromagnetic state of UGe$_2$\cite{saxena}. The value of $T_{sc}$ is 0.8 K at 1.2 GPa. URhGe and UCoGe show the superconducting transition at $T_{sc}$ = 0.2 and 0.7 K, respectively, at ambient pressure\cite{aoki,huy}. The characteristic temperature of the electronic state in Yb systems is lower than those in Ce and U systems due to the smaller band width of $4f$-band as mentioned before. If the superconductivity existed in YbCu$_2$Si$_2$, the transition temperature would be very low. This may be a reason why the heavy fermion superconductivity of the 4$f$ electrons is elusive in Yb systems. Also, spatial phase separation in YbCu$_2$Si$_2$ may be harmful for the appearance of the superconductivity. 

 The present study shows convincing evidence of ferromagnetism in the pressure-induced phase of YbCu$_2$Si$_2$ from dc magnetization measurements. Ferromagnetism has been found in a number of Yb compounds such as YbRhSb, YbInNi$_4$, and YbNiSn at ambient pressure\cite{muro,sarrao,kasaya}, YbInCu$_4$ and YbIr$_2$Si$_2$ at high pressure\cite{mito1,yuan}. On the other hand, there are only a few cerium based compounds such as CeRh$_3$B$_2$ and CeAg which show a ferromagnetic ground state\cite{dhar,cornelius}. The origin of this difference is an interesting question. The hierarchies of the energy scales of $T_{\rm K}$, ${\Delta}_{4f}$ and ${\Delta}_{CEF}$ in the Ce and Yb system are different as mentioned before\cite{harima,flouquet2,miyake}. This leads to the lager change of valence of Yb-ions from the non-magnetic 2+ to magnetic 3+ in real lattices, compared with that of the Ce ions. The resonant inelastic x-ray scattering experiment shows a wider valence change in YbCu$_2$Si$_2$, compared with that in its Ce-couterpart CeCu$_2$Si$_2$\cite{fernandez2}. The valence transition or instability of the Yb ion has been detected by the X-ray absorption or emission spectroscopy in YbAgCu$_4$\cite{nakamura}, YbInCu$_4$\cite{matsuda2}, and YbCu$_{5-x}$Al$_x$\cite{yamaoka}. Recently, new aspects in strongly correlated electron system originating from valence fluctuation of the rare earth ion have been theoretically discussed\cite{watanabe1,watanabe2}. Anomalous physical properties in $\beta$-YbAlB$_4$ and YbRh$_2$Si$_2$ have been re-considered from this point of view\cite{watanabe2}. The theoretical study also shows a simultaneous divergence of the valence susceptibility and the uniform spin susceptibility at the quantum critical point of the valence transition under a magnetic field. This strengthens a ferromagnetic tendency in Yb systems under finite magnetic field. Careful future theoretical study is necessary for a realization of the ferromagnetism under zero magnetic field\cite{watanabe3}. For the experimental point of view, comprehensive studies on the Yb systems should be done to reveal the valence state of the Yb ions in the wide temperature, magnetic field and pressure regions.

   \section{Conclusion}
 In conclusion, dc magnetic measurements have been done to study the magnetic property of the pressure-induced phase in YbCu$_2$Si$_2$ with a miniature ceramic anvil high pressure cell. The ferromagnetic ordered state is confirmed from the observation of the dc spontaneous magnetization. The pressure dependences of the ferromagnetic transition temperature $T_{\rm C}$ and the spontaneous magnetic moment ${\mu}_0$ at 2.0 K have been determined. The value of ${\mu}_0$ in the present macroscopic measurement is less than half of that determined via M\"{o}ssbauer experiment, which may be attributed to spatial phase separation between the ferromagnetic and paramagnetic phases. Peculiar features in the pressure-induced ferromagnetic state are discussed in comparison with cerium compounds. The effect of pressure on the magnetization with magnetic field along the magnetic easy $c$-axis is much larger than for field along the hard $a$-axis in the tetragonal structure. The pressure-induced phase in YbCu$_2$Si$_2$ has strong Ising-type uniaxial anisotropy, consistent with the two crystal electric field (CEF) models proposed for YbCu$_2$Si$_2$.

   \section{Acknowledgments}
We would like to thank the anonymous reviewers for their valuable comments and suggestions to improve the quality of the paper. We also acknowledge Prof. S. Watanabe for critical reading of this paper and giving us useful comments. This work was supported by a Grant-in-Aid for Scientific Research on Innovative Areas ``Heavy Electrons (Nos. 20102002 and 23102726), Scientific Research S (No. 20224015), A(No. 23246174), C (Nos. 21540373, 22540378 and 25400386), and Young Scientists (B) (No. 22740241) from the Ministry of Education, Culture, Sports, Science and Technology (MEXT) and Japan Society of the Promotion of Science (JSPS).

\bibliography{apssamp}

\end{document}